\documentclass[11pt,a4paper]{article}
\usepackage[utf8]{inputenc}
\usepackage{amsmath}
\usepackage{amsfonts}
\usepackage{hyperref}
\usepackage{amssymb}
\usepackage{makeidx}
\usepackage{cite}
\usepackage{bm}
\usepackage{geometry}
\usepackage{doc}

\usepackage{url}

\usepackage{graphicx}
\usepackage{epstopdf}
\begin{document}
\begin{titlepage}

~~\\

\vspace*{0cm}
    \begin{Large}
    \begin{bf}
       \begin{center}
         {Entanglement entropy and kinematic space in BCFT and RG flow}
       \end{center}
    \end{bf}   
    \end{Large}
\vspace{0.7cm}
\begin{center} 
Samrat Bhowmick$^b$\footnote
            {
e-mail address : 
tpsb5@iacs.res.in},
Suchetan Das$^{a}$\footnote
            {
e-mail address : 
suchetan.das@rkmvu.ac.in},
   
Bobby Ezhuthachan$^{a}$\footnote
            {
e-mail address : 
bobby.ezhuthachan@rkmvu.ac.in}\\

\vspace{0.7cm}
$^{a}$ {\it Ramakrishna Mission Vivekananda Education and Research Institute, Belur Math, Howrah-711202, West Bengal, India}\\
$^{b}$ {\it Department of Theoretical Physics,\\
Indian Association for the Cultivation of Science, Kolkata 700 032.  India.}\\
\end{center}

\vspace{0.7cm}

\begin{abstract}
The relation between kinematic space metric and entanglement entropy provides us with a differential equation for entanglement entropy in two dimension. For BCFT on upper half plane we solve this equation to obtain an expression for entanglement entropy consistent with known results in the literature. We also discuss how this relation can be used to recast the RG flow, under relevant deformations of a CFT, as a flow in the space of kinematic space metrics. 
\end{abstract}
\end{titlepage}
\pagenumbering{arabic}

\tableofcontents

%\tableofcontents

%%%%%%%%%%%%%%%%%%%%%%%%%%%%%%%%%%%%%%%%%%%%%%%%%%%%%%%%%%%%%%%
\section{Introduction}
In \cite{Czech:2015qta}, \cite{Czech:2015kbp} and \cite{Czech:2016xec}, the notion of the ``kinematic space" (k-space) as an intermediary between 
the AdS and CFT descriptions was introduced and developed further. In the CFT$_2$\footnote{In \cite{Czech:2016xec}, the k-space was discussed for CFT$_d$, 
but in this note our entire discussion is restricted to 2d CFT.}, the kinematic space is the space of pair of ordered space-like points or equivalently 
space of time-like points or space of causal diamonds\footnote{See also \cite{deBoer:2016pqk}.}. Objects like OPE
blocks, which are naturally a function of two points and have nice conformal transformation properties, are the simplest fields living on this k-space \footnote{This construction has been generalized to the ``boundary operators” that appear in the expansion of
the one point function of operators in the BCFT in \cite{Karch:2017fuh}, as well as for conical defects in \cite{Cresswell:2017mbk}.}.  

From the AdS$_3$ side, the k-space is the space of space-like geodesics which end on the boundary. Since such geodesics can also be parametrized by a pair of points on the boundary, this is again a space of ordered pair of points on the boundary of AdS. The fields on this geodesic space are the geodesic integral of scalar fields on AdS$_3$. The AdS/CFT dictionary [\cite{Maldacena:1997re} - \cite{Aharony:1999ti}], can then be recast as the statement of the identification of both these k-spaces as well as the fields(ie the OPE blocks with the geodesic integral of scalar fields in AdS). 
This gives a gauge invariant version of the AdS/CFT correspondence. 

One nice feature of the k-space formalism is that several known results can be recast/re-derived nicely in this language.  
\begin{itemize}
\item Identifying OPE blocks with Geodesic operators naturally leads to the known relation of geodesic Witten diagrams and conformal blocks 
\cite{Hijano:2015rla},\cite{Hijano:2015zsa},\cite{Hijano:2015qja}.
\item  Some new and some previously known results \cite{Casini:2011kv} regarding appearance of modular Hamiltonian in OPE of scalar operators.
\item By taking the inverse radon transform, the reconstruction of free bulk scalar fields in terms of boundary operators \cite{Hamilton:2006az} ,\cite{Hamilton:2005ju} has been re-derived also. Since then, some work has gone into extending this to include interactions \cite{Guica:2016pid}. (See also \cite{18.1}.) 
\item Re-deriving connections between linearised Einstein's equation \cite{Czech:2016tqr} from Entanglement,  previously discussed in [\cite{Lashkari:2013koa} - \cite{VanRaamsdonk:2016exw}], in this formalism.  
\end{itemize}

For the vacuum in CFT, the k-space metric is fixed completely by demanding that it be invariant under conformal transformation of the two points of the CFT \cite{Czech:2016xec}\footnote{This is consistent with the metric obtained via equation(\ref{ees}).}. In $CFT_{2}$, more generally, the metric on k-space is taken to be the second derivative of the entanglement entropy(EE) of a strip, with respect to the two endpoints\cite{Czech:2015qta}, \cite{Czech:2016xec}. On a single time slice, with end points(u, v) of the strip, the metric is given by: 
\begin{equation}\label{ees}
ds^{2}_{kin} = \frac{\partial^2 S}{\partial u\partial v} du dv
\end{equation}

One motivation for this is that, as shown in \cite{Czech:2015qta}, this leads to the correct formula for the expression of the differential entropy [\cite{Balasubramanian:2013lsa} - \cite{Czech:2014ppa}] in holographic theories\footnote{In holographic CFT$_{2}$, the differential entropy gives the length of closed curves in the bulk AdS$_{3}$, using the result of \cite{Ryu:2006bv}.}. 

In the literature, equation (\ref{ees}), has been used to compute the metric of the k-space for some excited states in the CFT. For instance, in \cite{Asplund:2016koz}, the authors computed the metric for the thermal state in a CFT. They used the metric so computed to identify the k-space geometry with that of the auxiliary de-Sitter space geometry discussed in \cite{deBoer:2015kda}.

In the first part of this note, we ask whether it is possible to use the relation between EE and kinematic space metric in the reverse direction? So that, instead of using (\ref{ees}) to get the form of the metric on k-space, we ask whether it's possible to interpret it as a differential equation for the EE and then solve it, using appropriate boundary conditions. In this approach, the RHS of (\ref{ees}) is interpreted as a source term for the partial differential equation, which has information of the system whose EE is being calculated. It would be a nice check on the entire approach, if this could be done. 

In section(\ref{sec:Kspace}), we show how to do this in the case of a single interval in the BCFT on the upper half plane (UHP). \footnote{Recently in \cite{Czech:2016nxc}, the kinematic space of defect, boundary and interface CFT has been studied using holography, in a different context from ours.} This serves as our prime example. The boundary condition we use, in this case is that in the 'small interval' limit($\frac{\Delta}{T}\rightarrow 0$) in which the interval length ($\Delta$) is much smaller than the distance of the interval from the boundary($T$), the EE must reduce to that of the full CFT vacuum result plus corrections. We also use another physically intuitive boundary condition that approaching this 'small interval' limit in two ways (ie: $\Delta\rightarrow 0$, $T$ finite and $\Delta$ finite and $T\rightarrow \infty$) must be physically equivalent. We show that the EE so obtained using this boundary condition, has a universal piece, which is completely fixed, and a piece with undetermined coefficients which depend on the explicit form of the source, and therefore upon the details of the particular BCFT theory being considered.

Our derivation of the universal piece matches with the known result in this case, which has been fixed previously using conformal invariance[\cite{Calabrese:2004eu} - \cite{Calabrese:2007mtj}]. 

In the appendix (\ref{kspace eg}), we provide two more applications of our method. The first corresponds to the case of the EE of an interval for the BCFT on a strip, while the other example is that of the EE of an interval  in any state corresponding to a descendant of the vacuum of the full CFT.  In both cases, following the methods introduced in section(\ref{sec:Kspace}), we show that we can reproduce the results known in the literature. For instance in the case of the strip, we show that in the limit when one of the endpoints of the interval lie on the boundary of the strip, we reproduce the universal term of the entanglement entropy \cite{Calabrese:2004eu}.  Similarly for the second example our result for the EE matches with the expression obtained in \cite{Holzhey:1994we},\cite{Sheikh-Jabbari:2016znt}.

In the second part of this note, we initiate a discussion on the kinematic space of CFT's deformed by relevant operators. The idea is that under RG flows, the EE changes and since the metric in kinematic space is the second derivative of the EE, it should be possible to use this, to reformulate the RG flow as a flow in the space of metrics on kinematic space. Our limited goal is to understand this change in the metric of the kinematic space under such flows. In particular it would be nice if we could write an equation of the following type:

\begin{equation}\label{flow}
E_{\mu\nu}[g, \Delta] = \lambda\frac{dg_{\mu\nu}}{d\lambda}
\end{equation}

Where $\lambda$ is the coupling constant of the relevant operator by which the CFT action ($I$) is being perturbed and $\Delta$ is the scaling dimension of  $\mathcal{O}$, the relevant deformation operator, which we take to be a primary scalar operator. ie: $I_{deformed} = I_{cft} +\lambda\int d^{d}x \mathcal{O}$. It would be nice if we could find a simple form for $E_{\mu\nu}$. We show that up to second order in the coupling, the RG flow can be expressed in the form of (\ref{flow}), with the explicit form of $E_{\mu\nu}$ given in eqn(\ref{secflow}). At higher orders, we have not been able to get a simple form for  the LHS of (\ref{flow}). It would depend very nontrivially on the dynamics of the theory as well as on the scaling dimension ($\Delta$) of the deforming operator.

\section{Kinematic space for CFT on Upper Half plane} \label{sec:Kspace}

Let us consider a two dimensional CFT on the upper half plane with complex coordinates, $z=x+ iy$ with $y\ge 0$. 

The symmetries of this upper half plane in terms of complex coordinates $(z=x+iy)$ are 
\begin{itemize}
\item $z' = z + c$, where $c$ is real;

\item $z' = \lambda z$, where $\lambda$ is real and positive;

\item $z' = \frac{z}{1+az}$, where a is real.
\end{itemize}

The invariant quantity under these transformations, which can be constructed out of two points is
\begin{equation}
\label{inv}
\eta = \left[\frac{(x_{1} - x_{2})^{2} + (y_{1} - y_{2})^{2}}{y_{1}y_{2}}\right]
\end{equation}
The most general form of the metric of kinematic space, invariant under these symmetries, is a function of $\eta$ and takes the following form
\begin{eqnarray}
ds^{2}_{kin} &=& f_{1}(\eta) \frac{dz_{1}\bar{dz_{1}}}{(z_{1}-\bar{z_{1}})^{2}} + f_{2}(\eta) 
\frac{dz_{2}\bar{dz_{2}}}{(z_{2}-\bar{z_{2}})^{2}} + f_{3}(\eta) \frac{dz_{1}dz_{2}}{(z_{1}-z_{2})^{2}} 
+ \bar{f_{3}}(\eta) \frac{\bar{dz_{1}}\bar{dz_{2}}}{(\bar{z_{1}}-\bar{z_{2}})^{2}} \nonumber \\
&+&f_{4}(\eta)\frac{dz_{1}\bar{dz_{2}}}{(z_{1}-\bar{z_{2}})^{2}} + \bar{f_{4}}(\eta)\frac{\bar{dz_{1}}dz_{2}}{(\bar{z_{1}}-z_{2})^{2}} 
\end{eqnarray}
The kinematic space is also invariant under the discrete reflection symmetry in the $x$ direction i.e $x\rightarrow -x$. We further impose the symmetry under the exchange of the two points i.e: $(z_{1},z_{2}) \rightarrow (z_{2},z_{1})$. 

Imposing these symmetries we can further restrict the form of the metric as,
\begin{eqnarray}\label{Kin}
ds^{2}_{kin} &=& g_{1}(\eta) \left[\frac{dz_{1}\bar{dz_{1}}}{(z_{1}-\bar{z_{1}})^{2}} + 
\frac{dz_{2}\bar{dz_{2}}}{(z_{2}-\bar{z_{2}})^{2}}\right] + g_{2}(\eta)\left[ \frac{dz_{1}dz_{2}}{(z_{1}-z_{2})^{2}} + 
\frac{\bar{dz_{1}}\bar{dz_{2}}}{(\bar{z_{1}}-\bar{z_{2}})^{2}}\right]\nonumber \\
&+&g_{3}(\eta)\left[\frac{dz_{1}\bar{dz_{2}}}{(z_{1}-\bar{z_{2}})^{2}} + \frac{\bar{dz_{1}}dz_{2}}{(\bar{z_{1}}-z_{2})^{2}}\right]
\end{eqnarray}

To go to the Lorentzian signature, We wick rotate and define $x=\tau$ and $y=i \sigma$, so that
$z=\tau+\sigma$ and $\bar{z} = \tau-\sigma$, where $\tau$ is a time like direction and the boundary is at $\sigma=0$.

Now consider  the kinematic space metric in a constant time slice. 
At constant time slice $\eta = \frac{(\sigma_{1} - \sigma_{2})^{2}}{\sigma_{1}\sigma_{2}}$ and 
the metric of the kinematic space can be written as 
\begin{equation}
\label{Kinconsttime}
ds^{2}_{kin} = g_{1}(\eta) \left[\frac{d\sigma_{1}d\sigma_{1}}{\sigma_{1}^{2}} + \frac{d\sigma_{2}d\sigma_{2}}{\sigma_{2}^{2}}\right] 
+ 2 \left[\frac{g_{2}(\eta)} {(\sigma_{1}-\sigma_{2})^{2}}+\frac{g_{3}(\eta)} {(\sigma_{1}+\sigma_{2})^{2}}\right]d\sigma_{1}d\sigma_{2} 
\end{equation}

However our assumption that kinematic space has a metric of the form given by equation (\ref{ees}), implies that for constant time slice the metric is given by
\begin{equation}
\label{adskin}
ds^{2} = \frac{\partial^{2}S_{EE}}{\partial \sigma_{1} \partial \sigma_{2}}d\sigma_{1} d\sigma_{2} =  ds^{2}_{kin}\big|_{\tau=\text{constant}}
\end{equation}
Comparing equations (\ref{Kinconsttime}) and (\ref{adskin}), one sees that $g_1(\eta)=0$.
Rest of the metric can be written as 
\begin{equation}
 ds^{2}_{kin} = \frac{g_{2}(\eta)+x^{-2}g_{3}(\eta)}{\Delta^{2}}d\sigma_{1}d\sigma_{2} 
\end{equation}
where $\Delta = (\sigma_{1} - \sigma_{2})$, $T = (\sigma_{1} + \sigma_{2})$ and $x = \frac{T}{\Delta}$.
$\eta$ is function of $x$ only, $\eta = \frac{4}{x^{2}-1}$.
Therefore, on a constant time slice kinematic space metric takes the form
\begin{equation}
\label{kinmet}
ds^{2}_{kin} = \frac{F(x^2)}{\Delta^{2}}d\sigma_{1}d\sigma_{2} 
\end{equation}

In the limit when the two points of the interval is far away from the boundary, one expects the boundary effects to be small, so that the entanglement entropy reduces to that of the CFT on the full plane. Mathematically, this limit is the following $\sigma_1,\sigma_2 \rightarrow \infty$, or $x \rightarrow \infty$.

\subsection{Entanglement entropy from kinematic space}\label{sec:EE}

From equations (\ref{adskin}) and (\ref{kinmet}), we get the following equation for the entanglement entropy.
\begin{equation}
\label{eeeqtn}
\frac{F(x)}{\Delta^2} = \frac{\partial^{2}S_{EE}}{\partial \sigma_{1} \partial \sigma_{2}}
\end{equation}
In this section, we try to solve this equation to obtain an expression for the entanglement entropy of a boundary CFT. 
We can rewrite this equation in terms of the variables $\Delta$ and $x$ to get
\begin{equation}
\label{main}
\frac{\partial^{2}S_{EE}(\Delta,x)}{\partial x^{2}} - \Delta^{2}\frac{\partial^{2}S_{EE}(\Delta,x)}{\partial \Delta^{2}} =F(x^{2})
\end{equation}
Observing that the right-hand side of equation (\ref{main}) is purely a function of $x^2$, motivates us to look for particular solutions of the following form
\begin{equation}
 \label{Sansatz}
 S_{EE} = S_1(x) + S_2(\Delta) 
\end{equation}
Plugging this back in (\ref{main}), we get
\begin{eqnarray}
\label{sep1}
\Delta^{2}\frac{d^{2}S_1(\Delta)}{d \Delta^{2}} -K &=& 0  \\
\label{sep2}
\frac{d^{2}S_2(x)}{d x^{2}} -K &=& F(x^{2}) 
\end{eqnarray}
%where $K$ is an arbitrary constant.
Where $K$ is an arbitrary constant. From the discussion below eqn (\ref{kinmet}), we expect that in the limit $x\rightarrow \infty$, $F(\infty)$ is finite and this motivates us to use the following series expansion for $F(x)$ in $\frac{1}{x^2}$ 
\begin{align}
F(x^{2}) = c_0 + \frac{c_{1}}{x^{2}} +\frac{c_{2}}{x^{4}} + \dots = \sum_{n=0}^{\infty} c_{n}x^{-2n}
\end{align}

Plugging this back, we can solve equations (\ref{sep1}) and (\ref{sep2}) for $S_{EE}$,
\begin{equation}
\label{sepsol}
S_{EE} = A\ln\Delta + B\ln T +C\Delta + ax + bx^{2} + (\frac{c'}{x^{2}} + \frac{c''}{x^{4}} + \dots) 
\end{equation}
where $A$, $B$, $C$ etc. are arbitrary constants and $A$ is related to $K$. 
One can check easily that (\ref{sepsol}) satisfies the equation (\ref{eeeqtn}). 
However, we know that given the wave equation with source, one can always add a term $f(\sigma_{1}) + g(\sigma_{2})$ to the solution because it satisfies the source free counterpart of the equation (\ref{eeeqtn}). Adding this general solution of the source free equation to the particular solution obtained through equations (\ref{Sansatz} -\ref{sepsol}), gives us the general solution the equation(\ref{main}), after imposing the appropriate boundary conditions, which we discuss below.

As we assume the symmetry $(\sigma_{1},\sigma_{2}) \rightarrow (\sigma_{2},\sigma_{1})$ in kinematic space, and therefore a symmetry of the entanglement entropy, in our case, $f$ and $g$ must be same, so that the full solution of $S_{EE}$ is of the form: 

\begin{equation}
\label{EESOL}
S_{EE} = A\ln\Delta + B\ln T +C\Delta + ax + bx^{2} + (\frac{c'}{x^{2}} + \frac{c''}{x^{4}} + \dots) + f(\sigma_1) +f(\sigma_2)
\end{equation}

\subsection {Constraining the form of $S_{EE}$ from various limits}
We will now try and constrain the form of $S_{EE}$ further, by comparing it with the form we expect it to take in well known limits.

%Now we will see entanglement entropy in various limiting case, which will help us to constrain the form of $S_{EE}$ further.

\subsubsection*{$\sigma_1 \to \sigma_2$ limit}
%The number of pure state in a quantum field theory is proportional to the spacelike volume of the system,
%in our $2$-dimensional case, the length of the system. While computing entanglement entropy, 
%one divides the region into two parts and trace out degrees of freedom of one part. So when the length of one part of the region shrink to zero, one trace out nothing and get a pure state. That is 
From the definition of the entanglement entropy it follows that when the length of the interval tends to zero,i.e $\sigma_{1} \rightarrow \sigma_{2}$\footnote{$\epsilon$ is a short distance scale, so that taking this limit means taking $\sigma_2 -\sigma_1 =\epsilon $.}, the entanglement entropy must be the entropy of total pure state and therefore vanishes. i.e
\begin{equation}
 \label{0length}
 \lim_{\sigma_1 \to \sigma_2} S_{EE} = 0 
\end{equation}
This limit implies that in(\ref{EESOL}), $f(\sigma) = -\frac{B}{2}\ln\frac{2\sigma}{\epsilon}$ and $a=b=0$.

\subsubsection*{``Far from the boundary'' limit}
When $\sigma_{1},\sigma_{2} \rightarrow \infty$ keeping the interval $(\sigma_{1}-\sigma_{2})$ finite, we reach a region where we expect the boundary effects to be negligible and therefore we should get the same expression for the entanglement entropy as in the full CFT. In the full CFT the entanglement entropy
is given by
\begin{equation}
S_{EE} = \frac{c}{3}\ln\frac{(\sigma_{1}-\sigma_{2})}{\epsilon} 
\end{equation}
This fixes $A=\frac{c}{3}$ and $C=0$. Here $\epsilon$ is introduced as an ultra violet cut-off.
Thus we have
\begin{equation}
\label{form}
S_{EE} = \frac{c}{3}\ln\frac{\Delta}{\epsilon} +B \ln\frac{T}{\epsilon} - 
\frac{B}{2}\ln\frac{4\sigma_{1}\sigma_{2}}{\epsilon^{2}} +  (\frac{c'}{x^{2}} + \frac{c''}{x^{4}} + \dots) + d 
\end{equation}
where $d$ is some arbitrary constant.

\subsubsection*{Further constraints}
We can further constrain the form of the entanglement entropy by the following physical consideration. The limit in which we expect the boundary effects to be negligible is the limit in which $\frac{\Delta}{T}<<1$. This limit can be approached in two ways.
%The limit in which we expect the boundary effects to be negligible is $\frac{\Delta}{T}<<1$. We use
%this fact to restrict the form of $S_{EE}$ further.
%This limit can be approached in two ways. 
\begin{itemize}
\item[1] $\Delta$ small and $T$ finite. 
\item[2] $\Delta$ finite and $T$ large. 
%The limit in which the length of the interval is finite, but the distance from the boundary is taken large. 
\end{itemize} 
We expect physically that both these limits to be the same. Mathematically this requirement translates to fixing $B=-A= -c/3$.
So as the final form of the entanglement entropy we find
\begin{equation}
\label{SEE-fin}
S_{EE} = \frac{c}{3}\ln\frac{\Delta}{\epsilon} - \frac{c}{3} \ln\frac{T}{\epsilon} + 
\frac{c}{6}\ln\frac{4\sigma_{1}\sigma_{2}}{\epsilon^{2}} +  (\frac{c'}{x^{2}} + \frac{c''}{x^{4}} + \dots) + d 
\end{equation}
%If there are scale invariance preserving boundary conditions, then $S_{EE}$ should also be a scale invariant quantity. 

The universal, theory independent log part is then completely fixed and agrees with the known result in literature. In the special case when the interval ends on the boundary of the UHP,is at the origin of the UHP, ie for say ($\sigma_{1}=0,\sigma_{2}=l$) or $\Delta=T$, (\ref{SEE-fin}) reduces to the result. 
\begin{equation}\label{SEE-fin2}
S_{EE} = \frac{c}{6}\ln\left(\frac{2l}{\epsilon}\right) + c'_{1}
\end{equation}
Which agrees with the result given in\cite{Calabrese:2004eu}.
 
The non universal part has undetermined coefficients, which would be theory dependent. This part contains a series expansion of $x^{2}$. We already noticed that the invariant quantity $\eta$ on the UHP is also a function of $x^{2}$. This $\eta$ is basically the cross ratio that appears while doubling the two point function on the UHP to its mirror. It is known that the entanglement entropy can be expressed in terms of the two point function of twist operators, on the UHP, and therefore should be a function of this cross ratio. Therefore our analysis gives the theory dependent part of entanglement entropy as a function of cross ratios as expected.

Finally we can compute the expression for the kinematic space 
metric from this entanglement entropy expression. It can be easily checked that in the limit 
$T\rightarrow \infty$ the expression for the kinematic space metric does reduce to the one given in \cite{Czech:2016xec}.

 To summarise, we inverted the relation between entanglement entropy and metric on k-space, to calculate the entanglement entropy for BCFT on UHP. We provide a few more examples in the appendix (\ref{kspace eg}). 

In BCFT, the universal part can be derived using conformal symmetry. The kinematic space formalism is a geometric way to encode the conformal symmetry. Therefore it is satisfying to see that we can reproduce the same result, from the geometric formulation of the conformal symmetry, as a solution of a PDE in the k-space- A language in which the replica trick or twist operators don't make a direct appearance.

\section{RG flows and kinematic space in deformed CFT's}\label{EEdeform}

While the k-space formalism provides a nice way to encode geometrically the kinematic conformal structure of CFT's, one could ask whether, beyond conformal kinematics, there are aspects of a CFT, which could be recast geometrically in this formalism\footnote{ A natural question is to ask whether the dynamics of the theory can be captured in the k-space language. See \cite{Karch:2017fuh} for some ideas in this direction.}. We end our note, by pointing out that the RG flow away from the CFT limit, may be recast as a flow in the space of kinematic space metrics\footnote{See \cite{Antonelli:2018qwz} for a discussion on RG flows and holographic integral geometry.}. This is because of the  relationship between the EE and the k-space metric, and the fact that under RG flows, the EE of a system changes. The general form of such a flow equation would be 

\begin{equation}\label{floweqn}
\Big[ \lambda \frac{d g_{\mu\nu}}{d\lambda} = E_{\mu\nu}\Big(g_{\mu\nu}, \Delta\Big) \Big]
\end{equation}

$\Delta$ is the scaling dimension of the deforming operator, and $g_{\mu\nu}$ is the k-space metric.

For a QFT, deformed from a CFT by a primary scalar operator $\mathcal{O}(x)$ of dimension $\Delta$, the action is given by
\begin{align}
I=I_{0}+\lambda\int d^{d}x \mathcal{O}(x)
\end{align}
where $I_{0}$ is the CFT action. 

Under a flow by such an deformation operator, the change in EE of the theory was studied in [\cite{Rosenhaus:2014zza}-\cite{Faulkner:2014jva}]. The formal expression for the change $\delta S$ was given by \cite{Rosenhaus:2014ula}
% the change in EE under such a flow is given by\cite{}
\begin{align}
\delta S= -<\mathcal{O}K>\lambda + \left(<\mathcal{O}\mathcal{O}K>-<\mathcal{O}\mathcal{O}>\right)\frac{(\lambda)^{2}}{2} + \dots \nonumber \\
= \sum_{n=1}(-1)^{n}\left(<K\underbrace{\mathcal{O}\dots\mathcal{O}}_{n}>- (n-1)<\underbrace{\mathcal{O}\dots\mathcal{O}}_{n}>\right) \frac{(\lambda)^{n}}{n!}
\end{align}
Where $K$ is the modular Hamiltonian, $<\dots>$ is vacuum expectation value and $\mathcal{O}$ refers to integral of $\mathcal{O}(x)$ over the entire manifold. 

We would like to apply these method to the case of 2d CFT's deformed by a relevant operator, to find the general dependence of the EE and therefore of the the k-space metric on $(\delta= v-u)$.  Order by order in ($\lambda$), the metric would be of the form: 
\begin{align}
g_{uv} \rightarrow g^{(0)}_{uv}+g^{(1)}_{uv}\lambda+g^{(2)}_{uv}\frac{\lambda^{2}}{2}+\dots
\end{align}
Here $g_{uv}$ is the non vanishing off-diagonal component of the two dimensional metric. 

We would then like to see if the corrected metric can be expressed as a solution of a flow equation of the type given in (\ref{floweqn}) .

Up to second order perturbation in $\lambda$, we show that the metric satisfies the equation 
\begin{equation}
\lambda\frac{d g^{(2)}_{\mu\nu}}{d\lambda} = \frac{2c_{0}}{(4-2\Delta)(3-2\Delta)-2}\left(R^{(2)}_{\mu\nu} - \frac{2}{c_{0}}g^{(2)}_{\mu\nu}\right)
\end{equation}
where $c_{0} = \frac{c}{3}$ and $c$ is the central charge of the CFT.
%We provide the details of the calculation in the next section. Our goal is to find only the dependence of the metric on $\delta$ and so we do not compute the integrals explicitly. We comment on this as well in the appendix. 
At higher orders, the metric correction is of the form:

\begin{equation}
g_{uv} = \frac{c_{0}}{\delta^{2}} + \frac{1}{\delta^2}\sum_{n=2}C_{n}\lambda^{n}\delta^{(2-\Delta)n}
\end{equation}

We now provide the details of the calculation.

\subsection{Details of the calculation of k-space metric under RG flow}

\begin{itemize}
\item \textit{Ricci tensor for the metric}
\end{itemize}
In any holographic theory, the form of the kinematic space metric in two dimension, on constant time slice, is related to the entanglement entropy as,
\begin{align}
ds^{2} = \frac{\partial^{2}S_{EE}(u,v)}{\partial u \partial v}dudv
\end{align}
Therefore only off-diagonal components of this 2-dimensional metric is non-vanishing.
\begin{align}
g_{\mu\nu}=
\begin{bmatrix}
0 & g_{uv} \\
g_{uv} & 0
\end{bmatrix}
\quad ; \quad g^{-1}_{\mu\nu}=
\begin{bmatrix}
0 & \frac{1}{g_{uv}} \\
\frac{1}{g_{uv}} & 0
\end{bmatrix}
\end{align}
The only non-zero components of Christoffel symbols are,
\begin{align}
\Gamma^{u}_{uu} = g^{uv}\partial_{u}g_{uv} \quad ; \quad \Gamma^{v}_{vv} = g^{uv}\partial_{v}g_{uv}
\end{align}
From this we can get the following components of Ricci tensors
\begin{align}\label{Ruv}
R_{uv}=-\partial_{v}\left(g^{uv}\partial_{u}g_{uv}\right); \quad R_{vu}=-\partial_{u}\left(g^{uv}\partial_{v}g_{uv}\right); \quad R_{uu}=R_{vv}=0
\end{align}
Note that, for any two dimensional metric containing only off-diagonal components $g_{uv}$, the Einstein tensor $G_{uv}$ vanishes identically as Ricci scalar $R$ is just $2R_{uv}g^{uv}$
\begin{align}
G_{uv}=R_{uv}-\frac{1}{2}g_{uv}R = 0
\end{align}

\begin{itemize}
\item \textbf{zeroth order metric}
\end{itemize}
The zeroth order part of the metric is the CFT k-space metric, which is already fixed
\begin{align}
g^{(0)}_{uv} = \frac{c_{0}}{(v-u)^{2}} \quad ; \quad g^{(0)uv}=\frac{(v-u)^{2}}{c_{0}}
\end{align}
where $c_{0} = \frac{c}{3}$ and $c$ is the central charge of the theory. The Ricci tensor for this metric is,
\begin{align}
R_{uv}^{(0)}=-\partial_{v}\left(g^{(0)uv}\partial_{u}g^{(0)}_{uv}\right) = \frac{2}{c_{0}}g^{(0)}_{uv} 
\end{align}
This is the constant curvature de-Sitter metric with Ricci scalar $R^{(0)} = \frac{4}{c_{0}}$. 

\begin{itemize}
\item \textbf{first order correction of metric}
\end{itemize}
The first order correction of entanglement entropy is $\delta S^{(1)}= -<\mathcal{O}K>\lambda$. In CFT$_{2}$, for a single interval, we have nice expression of modular Hamiltonian $K$ [\cite{Casini:2011kv},\cite{Cardy:2016fqc}] in terms of integral over local integrand proportional to stress-energy tensor. More explicitly,
\begin{align}
\delta S^{(1)}= -\int d\omega_{1}d\bar{\omega}_{1}\int_{u}^{v}dx\frac{(x-u)(v-x)}{(v-u)}<T_{00}(x)\mathcal{O}(\omega_{1}, \bar{\omega}_{1})>
\end{align}
But for any scalar primary operator in CFT, the two point function $<T_{00}(x)\mathcal{O}(x_{1})>$ vanishes. Therefore $\delta S^{(1)}$, and hence $g^{(1)}_{uv}$ does not contribute to the flow. The first non-vanishing contribution comes from the next order correction.
%{\it Thus (\ref{flow-zero}) continues to hold at this order also}.

\begin{itemize}
\item \textbf{second order correction of metric}
\end{itemize}
The second order change in EE is,
\begin{eqnarray}
\delta S^{(2)}= &&\int d\omega_{1}d\bar{\omega}_{1}\int d\omega_{2}d\bar{\omega}_{2}\int_{u}^{v}dx\frac{(x-u)(v-x)}{(v-u)}<\mathcal{O}(\omega_{1},\bar{\omega}_{1})\mathcal{O}(\omega_{2},\bar{\omega}_{2})T_{00}(x)> \nonumber \\
&&- \int d\omega_{1}d\bar{\omega}_{1}\int d\omega_{2}d\bar{\omega}_{2}<\mathcal{O}(\omega_{1},\bar{\omega}_{1})\mathcal{O}(\omega_{2},\bar{\omega}_{2})>
\end{eqnarray}
As we are ultimately interested to calculate the contribution to the metric $\frac{\partial^{2}\delta S^{(2)}_{EE}(u,v)}{\partial u \partial v}$, the contribution from the second part of two point function of $\mathcal{O}$ vanishes. Therefore we get,
\begin{align}
g^{(2)}_{uv} \sim (\lambda)^{2}\int_{u}^{v}dx\frac{(x-u)(v-x)}{(v-u)^{3}}\int d^{2}\omega_{1}\int d^{2}\omega_ {2}<\mathcal{O}(\omega_{1},\bar{\omega}_{1})\mathcal{O}(\omega_{2},\bar{\omega}_{2})T_{00}(x)> \nonumber \\
= \frac{\Delta}{2} \int_{u}^{v}dx\frac{(x-u)(v-x)}{(v-u)^{3}}\int d^{2}\omega_{1}\int d^{2}\omega_{2}\Big[\frac{(\omega_1 -\omega_2)^2}{|(\omega_1 -\omega_2)|^{2\Delta}(x-\omega_{1})^{2}(x-\omega_{2})^{2}}+c.c\Big ]
\end{align}
%where $c'$ is some constant. 
{\it A few observations about the integrals} 
\begin{itemize}
\item 
Naively, the integrals over $\omega_1$ and $\omega_2$ can take various values. For instance, if we absorb $x$ into $\omega_i$($i=$1,2) by shifting $\omega_{i}\rightarrow w_{i}-x$, then the integrals are independent of $x$. On the other hand if we scale the $\omega_{i}\rightarrow x\omega_{i} $, then integral is proportional to $x^{-(2\Delta-2)}$. This is because the integrals are not well defined, due to both infrared and ultraviolet divergences. We therefore have to regulate the integrals. 

\item
We expect the answer to be a function of $\delta =(v-u)$. We first make this manifest by expressing everything in terms of $\delta$. To do this we shift $x$ by $y=x-u$, to get:
\begin{equation}
g^{(2)}_{uv} = \frac{\Delta}{2} \int_{0}^{\delta}dx\frac{y(\delta-y)}{\delta^{3}}\int \frac{d^{2}\omega_{1} d^{2}\omega_{2}}{|(\omega_1 -\omega_2)|^{2\Delta}}\Big[\frac{(\omega_1 -\omega_2)^2}{(y-\omega_{1})^{2}(y-\omega_{2})^{2}}+c.c\Big ]
\end{equation}

\item 
Naively the integral is independent of $y$, because it can be absorbed into a shift of $(T=\omega_{1}+\omega_{2})$. We first regulate this $T$ integral by introducing a Gaussian term $(e^{-s(\frac{|T|^2}{y^2})})$, which would damp the large $T$ behaviour of the integral. Then take the $(s\rightarrow 0)$ limit. 

\item We now pull out the $y$ factors explicitly, by scaling $\omega_{i}\rightarrow y\omega_{i}$. The remaining integral, has to be further regulated, but since we are interested in the $y$ dependence, we simply denote this regulated constant as $c_2$. 

\end{itemize}

\begin{equation}\label{secondorder}
g^{(2)}_{uv} = \frac{c'\Delta}{2} \int_{0}^{\delta}dx\frac{y(\delta-y)y^{2-2\Delta}}{\delta^{3}} = c_{2}(\Delta)\frac{\Delta}{2(4-2\Delta)(5-2\Delta)}\delta^{2-2\Delta}
\end{equation}

This last integral is well defined only for relevant perturbations $\Delta<2$. For $\Delta >2$ the integral blows up at the lower limit. As expected for marginal operators, $\Delta\rightarrow 2$, the correction is same as the zeroth order $g^{2}\sim \delta^{-2}$. In \cite{Faulkner:2014jva}, the author computes the finite second order change of EE in any dimension, under such a deformation and for 2D the length of the interval has the same power law behaviour, which we get in (\ref{secondorder}). But we do not explicitly compute the constant term $c_{2}$.

Now we would like to see, whether up to second order, we can write an equation of the form (\ref{floweqn}) for such flow. Let us consider the following Ricci flow equation as our ansatz for geometric flow equation:
\begin{equation}\label{ansatz}
\lambda \frac{d g_{uv}}{d\lambda} =  a g_{uv}+ b R_{uv}
\end{equation}
From the zeroth order we can easily see that $a=-\frac{2b}{c_{0}}$. From the second order of the equation we can fix $a,b$. From (\ref{Ruv}) and (\ref{secondorder}) one can compute second order correction of Ricci tensor:
\begin{align}
R^{(2)}_{uv} = \frac{c_{2}(4-2\Delta)(3-2\Delta)}{c_{0}}(v-u)^{2-2\Delta}
\end{align} 
Putting it back into (\ref{ansatz}) and solving it, we get $a=\frac{4}{(4-2\Delta)(3-2\Delta)-2}$. Therefore the flow equation becomes,
\begin{align}\label{secflow}
\lambda\frac{d g_{uv}}{d\lambda} = \frac{2c_{0}}{(4-2\Delta)(3-2\Delta)-2}\left(R_{uv} - \frac{2}{c_{0}}g_{uv}\right)
\end{align}

\begin{itemize}
\item \textbf{third order correction of metric}
\end{itemize}
The third order correction of the metric is,
\begin{align}
g^{(3)}_{uv} \sim -\int_{u}^{v}dx\frac{(x-u)(v-x)}{(v-u)^{3}}\int d^{2}\omega_{1}\int d^2\omega_{2}\int d^2\omega_{3}<\mathcal{O}(\omega_{1},\bar{\omega}_{1})\mathcal{O}(\omega_{2},\bar{\omega}_{2})\mathcal{O}(\omega_{3},\bar{\omega}_{3})T_{00}(x)>
\end{align}
The four point function of $<\mathcal{O}(\omega_{1},\bar{\omega}_{1})\mathcal{O}(\omega_{2},\bar{\omega}_{2})\mathcal{O}(\omega_{3},\bar{\omega}_{3})T_{00}(x)>$ can be calculated from conformal Ward identity in which it can be recast as some linear combination of three point function $<\mathcal{O}(\omega_{1},\bar{\omega}_{1})\mathcal{O}(\omega_{2},\bar{\omega}_{2})\mathcal{O}(\omega_{3},\bar{\omega}_{3})>$ and sum of regular terms. Ignoring the regular part we get,

\begin{eqnarray}
&&<T(x)\prod^{3}_{i=1}\mathcal{O}(\omega_{i},\bar{\omega}_{i})> = \sum_{i=1}^{3}\left[\frac{1}{x-\omega_{i}}\partial_{\omega_{i}} + \frac{\Delta/2}{(x-\omega_{i})^{2}}\right]<\mathcal{O}(\omega_{1},\bar{\omega}_{1})\mathcal{O}(\omega_{2},\bar{\omega}_{2})\mathcal{O}(\omega_{3},\bar{\omega}_{3})>  + c.c \nonumber \\
&=& \frac{\Delta C_{\mathcal{O}\mathcal{O}\mathcal{O}}}{2|\omega_{12}|^{\Delta}||\omega_{23}|^{\Delta}|||\omega_{31}|^{\Delta}|}\Big[\frac{\omega^{2}_{12}}{(z-\omega_{1})^2(z-\omega_{2})^{2}} +\frac{\omega^{2}_{23}}{(z-\omega_{2})^2(z-\omega_{3})^{2}}+\frac{\omega^{2}_{31}}{(z-\omega_{1})^2(z-\omega_{3})^{2}}+c.c\Big]\nonumber
%= c_{123}\sum_{i=1}^{3}\left[\frac{1}{x-x_{i}}\partial_{x_{i}} + \frac{h_{i}}{(x-x_{i})^{2}}\right]\frac{1}{(x_{1}-x_{2})^{\Delta}(x_{2}-x_{3})^{\Delta}(x_{3}-x_{1})^{\Delta}} \nonumber \\
%= c_{123}\sum_{i=1}^{3}\partial_{x_{i}}\left(\frac{1}{(x-x_{i})(x_{1}-x_{2})^{\Delta}(x_{2}-x_{3})^{\Delta}(x_{3}-x_{1})^{\Delta}}\right) \nonumber \\
%+ c_{123}\left(\frac{\Delta}{2}-1\right)\sum_{i=1}^{3} \left(\frac{1}{(x-x_{i})^{2}(x_{1}-x_{2})^{\Delta}(x_{2}-x_{3})^{\Delta}(x_{3}-x_{1})^{\Delta}}\right)
\end{eqnarray}
Following the steps as before, we get:
\begin{equation}
g^{(3)}_{uv}= \frac{c_3\Delta C_{\mathcal{O}\mathcal{O}\mathcal{O}}}{2}\int^{\delta}_{0}dy\frac{y(\delta-y)y^{4-3\Delta}}{\delta^3} =c_{3}(\Delta)\frac{\delta^{4-3\Delta}}{(6-3\Delta)(7-3\Delta)}
\end{equation}
Here $c_3$ is again the regulated integral as before. Also as before, this integral is well defined only for relevant perturbations $\Delta<2$.  Again, if we compute $R^{(3)}_{uv}$ and put it back into the equation (\ref{secflow}) , we see it does not satisfy the equation. 

\begin{itemize}
\item \textbf{fourth and higher order correction of metric}
\end{itemize}
At the fourth order, Ward identity tells us that we will need to compute a four point function of $\mathcal{O}$.
\begin{equation}
<T(x)X(\omega,\bar{\omega})> = \sum^{4}_{i=1}\Big[\frac{1}{(x-\omega_{i})}\partial_{i}<X> + \frac{\Delta/2}{(x-\omega_{i})^2}<X> +c.c\Big]
\end{equation}

Here $X= \mathcal{O}(\omega_{1},\bar{\omega}_{1})\mathcal{O}(\omega_{2},\bar{\omega}_{2})\mathcal{O}(\omega_{3},\bar{\omega}_{3})\mathcal{O}(\omega_{4},\bar{\omega}_{4})$. 
\begin{equation}
<X> = \frac{F(z,\bar{z})}{|(\omega_{1}-\omega_2)|^{\alpha_{12}}|(\omega_{1}-\omega_3)|^{\alpha_{13}}|(\omega_{1}-\omega_4)|^{\alpha_{14}}|(\omega_{2}-\omega_3)|^{\alpha_{23}}|(\omega_{4}-\omega_2)|^{\alpha_{24}}|(\omega_{3}-\omega_4)|^{\alpha_{34}}}
\end{equation}
where $\sum \alpha_{ij}=4\Delta$ and $z = \frac{(\omega_{1}-\omega_{2})(\omega_{3}-\omega_{4})}{(\omega_{1}-\omega_{3})(\omega_{2}-\omega_{4})}$ is the cross ratio. $F(z,\bar{z})$ is a function of cross ratio which depends on the dynamics and spectrum of the theory. 

One can now follow the same steps as previously done to regulate and separate out the $y$ dependence from the integrals. While the integral contains an undetermined function $F$, but it does not affect the scaling arguments and the final $y$ integral takes the form:
\begin{align}
g^{(4)}_{uv} = c_{4}(\Delta) \int_{0}^{\delta}dy\frac{y(\delta-y)}{\delta^{3}}y^{6-4\Delta} = c_{4}(\Delta) \frac{y^{6-4\Delta}}{(8-4\Delta)(9-4\Delta)}
\end{align}
The difference from the lower order is that at this order, the $c_{4}(\Delta)$ contains the information about the full spectrum of operators  in the CFT and hence has to be determined separately for each theory.

The story is the same at higher orders and we can use the same method as  for four point function. In particular the scaling of n-point function tells us that the contribution of $y$ coming from the integrals of $<T(x)\Big(\prod^{n}_{i=1}\mathcal{O}(\omega_{i},\bar{\omega}_{i})\Big)>$ is just $y^{n-n\Delta-2}$. Thus,
\begin{align}
g^{(n)}_{uv} \sim \int_{0}^{\delta}dy\frac{y(\delta-y)}{(\delta)^{3}}y^{2n-2-n\Delta} \sim \frac{\delta^{2n-2-n\Delta}}{(2n-n\Delta)(2n+1-n\Delta)}
\end{align}
Therefore the general form of the metric is,
\begin{align}\label{main}
g_{uv} = \frac{c_{0}}{\delta^{2}} + \frac{1}{\delta^2}\sum_{n=2}c_{n}\lambda^{n}\delta^{(2-\Delta)n}
\end{align}
Where, the contributions of the regulated integrals and other factors are absorbed into the constant $c_n$.

Its difficult to find an explicit form of (\ref{floweqn}) for this general metric (\ref{main}). We have not tried to do so in this note. It would be nice if starting from the bulk, a similar flow equation could be derived in the holographic/AdS k-space  ie "the space of geodesics". We hope to return to these questions in the future.

\vspace*{1ex}
\noindent{\bf Acknowledgment:} 
We would like to thank James Sully for useful discussions, and for his suggestions and comments on a preliminary version of this note. We would also like to thank Dileep Jatkar, Arnab Kundu, and Koushik Ray for useful discussions. The research work of Suchetan Das is supported by a fellowship from CSIR.

\appendix
\section{EE from k-space: Two examples}\label{kspace eg}
In this appendix, we provide some more examples in which we use the methods introduced in section (\ref{sec:Kspace}) to compute the EE.  

\subsection{EE for single interval on finite strip from k-space}
We set up the calculation of the entanglement entropy for any arbitrary two points $x_{1}$, $x_{2}$ in the finite strip of length $L$. The kinematic space metric for this case can be obtained by a conformal transformation ($z = e^{\frac{\pi}{L}\omega}$, $\omega = t + ix$)from the upper half plane k-space metric(\ref{Kin}), yielding the following expression:

\begin{align}
ds^{2}= \frac{4 \pi^{2}}{L^{2}}[\frac{g_{2}(\eta)}{\sinh^{2}\frac{\pi(\omega_{1} - \omega_{2})}{L}}d\omega_{1}d\omega_{2} + \frac{g_{2}(\eta)}{\sinh^{2}\frac{\pi(\bar{\omega_{1}} - \bar{\omega_{2}})}{L}}d\bar{\omega_{1}}d\bar{\omega_{2}} + \frac{g_{3}(\eta)}{\sinh^{2}\frac{\pi(\omega_{1} - \bar{\omega_{2}})}{L}}d\omega_{1}d\bar{\omega_{2}} + \frac{g_{2}(\eta)}{\sinh^{2}\frac{\pi(\bar{\omega_{1}} - \omega_{2})}{L}}d\bar{\omega_{1}}d\omega_{2} ]
\end{align}

At the constant time slice, the k-space metric simplifies into the following
\begin{align}\label{finite}
ds^{2} =\frac{2 \pi^{2}}{L^{2}}[\frac{g_{2}(\eta)}{\sin^{2}(\frac{\pi(x_{1} - x_{2})}{2L})(1-\sin^{2}(\frac{\pi(x_{1} - x_{2})}{2L}))} + \frac{g_{3}(\eta)}{\sin^{2}(\frac{\pi(x_{1} + x_{2})}{2L})(1-\sin^{2}(\frac{\pi(x_{1} + x_{2})}{2L}))}]dx_{1}dx_{2} 
\end{align}

With $\eta$ at constant time slice, given by:
\begin{align}\label{eta}
\eta = \frac{2\sin^{2}\frac{\pi}{2L}(x_{1} - x_{2})}{\sin^{2}\frac{\pi}{2L}(x_{1} + x_{2}) - \sin^{2}\frac{\pi}{2L}(x_{1} - x_{2})} = \frac{2}{\frac{\sin^{2}\frac{\pi}{2L}(x_{1} + x_{2})}{\sin^{2}\frac{\pi}{2L}(x_{1} - x_{2})} - 1}
\end{align}

Following the arguments of section(\ref{sec:Kspace}), we solve the k-space equation to get the EE for this case. Let us define $\beta = \sin\frac{\pi\Delta}{2L}$ and $\alpha = \sin\frac{\pi T}{2L}$ where $\Delta = (x_{1} - x_{2})$, $T = (x_{1} + x_{2})$. In terms of $\alpha$ and $\beta$ we can re-express the kinematic space equation as follows:
\begin{align}\label{SStrip}
(1-\alpha^{2})\frac{\partial^{2}S_{EE}}{\partial \alpha^{2}} - \alpha \frac{\partial S_{EE}}{\partial \alpha} - (1-\beta^{2})\frac{\partial^{2}S_{EE}}{\partial \beta^{2}} + \beta \frac{\partial S_{EE}}{\partial \beta} = 8[\frac{g_{2}(\eta)}{\beta^{2}(1-\beta^{2})} +\frac{g_{3}(\eta)}{\alpha^{2}(1-\alpha^{2})}]
\end{align}
\begin{itemize}
\item \textbf{A special case}

Instead of solving the above equation in full generality, we will try to solve it for a special form of the source, namely we take $g_2$ and $g_3$, to be constants, for which case, the metric takes the following form:
%To attack this special case we chose a clever way of finding entanglement entropy for special kind of source and later we argue that this must be the special case we mentioned above.
%Let us extract a special kind of source from \ref{finite} which is
\begin{align}\label{spl}
ds^{2} =\frac{a}{\sin^{2}(\frac{\pi \Delta}{L})} + \frac{b}{\sin^{2}(\frac{\pi T}{L})}dx_{1}dx_{2} 
\end{align}
where $a,b$ are constants. For this form of the kinematic space metric, we solve the differential equation for entanglement entropy using the separation of variable ansatz $S_{EE} = S_{1}(\Delta) + S_{2}(T)$ and find the following expression
\begin{align}\label{EESTRIP}
S(x_{1},x_{2}) = c'\ln(k'\sin \frac{\pi \Delta}{L}) + b'\Delta^{2} + d'\Delta + \nonumber \\
 c''\ln(k''\sin \frac{\pi T}{L}) - b'T^{2} + d''T + f(\Delta+T) + f(T-\Delta) +a
\end{align}

For the case when one of the points is on the boundary of the strip, the expression simplifies to the following form.
%Now we will see for which case one can get this type of source for kinematic space. From \ref{eta} it is clear that $\eta$ must be a function of $\frac{\sin^{2}\frac{\pi}{2L}T}{\sin^{2}\frac{\pi}{2L}\Delta}$. Therefore to get constant $g_{2}(\eta),g_{3}(\eta)$ in \ref{finite} for generic $\Delta$ and $T$ one must have $\Delta = T$. In such case EE must be the function of $x_{1}$ only. Therefore we have 
\begin{align}
S(x_{1}) = g\ln(k\sin \frac{\pi x_{1}}{L}) + F(x_{1}) + a'
\end{align}
Again the zero interval limit for EE fixes $F(x)$ to be zero. Also in the limit $x_{1}<<L$, we expect the entanglement entropy to be the same as that obtained for the EE on the UHP(\ref{SEE-fin2}). This fixes $g$ and $k$. Thus, we get:
\begin{align}
S(x_{1}) = \frac{c}{6}\ln(\frac{2L}{\pi\epsilon}\sin \frac{\pi x_{1}}{L}) + a
\end{align}

Note that we assumed a particular form of the source to derive this last result.  However we expect this to hold for any  form of the source term. This is because, to derive the expression for $S(x_1, 0)$, we need to solve the eqn(\ref{SStrip}) in the neighbourhood of the the $\Delta =T$ slice, where we can take $g_{2}(\eta)$  and $g_{3}(\eta)$ to be constants. Thus on that slice, any source term will take the form given in (\ref{spl}) and therefore the analysis presented above should hold.  

\end{itemize}

\subsection{EE for vacuum descendant states}
In this section we will reproduce known result [\cite{Holzhey:1994we},\cite{Sheikh-Jabbari:2016znt}]  for entanglement entropy for a vacuum descendant state in the kinematic space language. Under an arbitrary local conformal transformation $z \rightarrow h(z)$, the metric for kinematic space on a constant time slice becomes,
\begin{align}\label{loc}
ds^{2}_{loc}\equiv \frac{\partial^{2}S^{(h)}(x_{1},x_{2})}{\partial x_{1}\partial x_{2}}dx_{1}dx_{2} =\frac{a dx_{1}dx_{2}}{(h(x_{1})-h(x_{2}))^{2}}\frac{dh(x_{1})}{dx_{1}}\frac{dh(x_{2})}{dx_{2}} 
\end{align}
Where $a$ is a constant not determined by symmetry and $S^{(h)}(x_{1},x_{2})$ is the EE of the  of interval $(x_{1},x_{2})$. Therefore the general solution for $S^{(h)}(x_{1},x_{2})$ is
\begin{align}
S^{(h)}(x_{1},x_{2}) = a\ln \left(\frac{h(x_{1})-h(x_{2})}{\epsilon}\right)+g(x_{1})+g(x_{2})
\end{align}
As before, we further constrain the above expression, by considering the pure state limit i.e $\lim_{x_{1}-x_{2} \rightarrow \epsilon}S =0$. This fixes $g(x_{1}) = -\frac{a}{2}\ln h'(x_{1})$. Again for the special case of $h(x_{1}) = x_{1}$ we would have $S = \frac{c}{3}\ln \left(\frac{x_{1}-x_{2}}{\epsilon}\right)$ which fixes $a$. Thus we end up with the following expression for  the EE.
\begin{align}
S^{(h)}(x_{1},x_{2}) = \frac{c}{3}\ln \left(\frac{h(x_{1})-h(x_{2})}{\epsilon}\right)-\frac{c}{6}\ln\left( h'(x_{1})h'(x_{2})\right)
\end{align}

.


\begin{thebibliography}{99}

\bibitem{Czech:2015qta} 
  B.~Czech, L.~Lamprou, S.~McCandlish and J.~Sully,
  ``Integral Geometry and Holography,''
  JHEP {\bf 1510}, 175 (2015)
  doi:10.1007/JHEP10(2015)175
  [arXiv:1505.05515 [hep-th]].
  %%CITATION = doi:10.1007/JHEP10(2015)175;%%
  %44 citations counted in INSPIRE as of 22 Feb 2017
  
%\cite{Czech:2015kbp}
\bibitem{Czech:2015kbp} 
  B.~Czech, L.~Lamprou, S.~McCandlish and J.~Sully,
  ``Tensor Networks from Kinematic Space,''
  JHEP {\bf 1607}, 100 (2016)
  doi:10.1007/JHEP07(2016)100
  [arXiv:1512.01548 [hep-th]].
  %%CITATION = doi:10.1007/JHEP07(2016)100;%%
  %20 citations counted in INSPIRE as of 22 Feb 2017
  
\bibitem{Czech:2016xec} 
  B.~Czech, L.~Lamprou, S.~McCandlish, B.~Mosk and J.~Sully,
  ``A Stereoscopic Look into the Bulk,''
  JHEP {\bf 1607}, 129 (2016)
  doi:10.1007/JHEP07(2016)129
  [arXiv:1604.03110 [hep-th]].
  %%CITATION = doi:10.1007/JHEP07(2016)129;%%
  %29 citations counted in INSPIRE as of 22 Feb 2017
%\cite{Czech:2015qta}
  
%\cite{deBoer:2016pqk}
\bibitem{deBoer:2016pqk} 
  J.~de Boer, F.~M.~Haehl, M.~P.~Heller and R.~C.~Myers,
  ``Entanglement, holography and causal diamonds,''
  JHEP {\bf 1608}, 162 (2016)
  doi:10.1007/JHEP08(2016)162
  [arXiv:1606.03307 [hep-th]].
  %%CITATION = doi:10.1007/JHEP08(2016)162;%%
  %17 citations counted in INSPIRE as of 23 Feb 2017  
  
%\cite{Hijano:2015zsa}

  %%CITATION = doi:10.1007/JHEP01(2016)146;%%
  %41 citations counted in INSPIRE as of 23 Feb 2017
  
 
%%%%%%%%%%%%%%%%%%%%%%%%%%%%%%%%%%%%%%%%%%%%%%%%%%%%%%%%%%%%%%%%%%%%%%%%%%
%%%%%%%%%%%%%%%%%%%%%%%%%%%%%%%%%%%%%%%%%%%%%%%%%%%%%%%%%%%%%%%%%%%%%%%%%%  
\bibitem{Maldacena:1997re}
J.~M.~Maldacena,
  ``The Large N limit of superconformal field theories and supergravity,''
  Int.\ J.\ Theor.\ Phys.\  {\bf 38}, 1113 (1999)
  [Adv.\ Theor.\ Math.\ Phys.\  {\bf 2}, 231 (1998)]
  doi:10.1023/A:1026654312961, 10.4310/ATMP.1998.v2.n2.a1
  [hep-th/9711200].
\bibitem{Gubser:1998bc}
S.~S.~Gubser, I.~R.~Klebanov and A.~M.~Polyakov,
  ``Gauge theory correlators from noncritical string theory,''
  Phys.\ Lett.\ B {\bf 428}, 105 (1998)
  doi:10.1016/S0370-2693(98)00377-3
  [hep-th/9802109].
\bibitem{Witten:1998qj}
E.~Witten,
  ``Anti-de Sitter space and holography,''
  Adv.\ Theor.\ Math.\ Phys.\  {\bf 2}, 253 (1998)
  doi:10.4310/ATMP.1998.v2.n2.a2
  [hep-th/9802150].
  \bibitem{Aharony:1999ti} 
  O.~Aharony, S.~S.~Gubser, J.~M.~Maldacena, H.~Ooguri and Y.~Oz,
  ``Large N field theories, string theory and gravity,''
  Phys.\ Rept.\  {\bf 323}, 183 (2000)
  doi:10.1016/S0370-1573(99)00083-6
  [hep-th/9905111].

%\cite{Ryu:2006bv}
\bibitem{Ryu:2006bv} 
  S.~Ryu and T.~Takayanagi,
  ``Holographic derivation of entanglement entropy from AdS/CFT,''
  Phys.\ Rev.\ Lett.\  {\bf 96}, 181602 (2006)
  doi:10.1103/PhysRevLett.96.181602
  [hep-th/0603001].
  %%CITATION = doi:10.1103/PhysRevLett.96.181602;%%
  %1179 citations counted in INSPIRE as of 05 Mar 2017
%\cite{Hijano:2015rla}
\bibitem{Hijano:2015rla} 
  E.~Hijano, P.~Kraus and R.~Snively,
  ``Worldline approach to semi-classical conformal blocks,''
  JHEP {\bf 1507}, 131 (2015)
  doi:10.1007/JHEP07(2015)131
  [arXiv:1501.02260 [hep-th]].
  %%CITATION = doi:10.1007/JHEP07(2015)131;%%
  
\bibitem{Hijano:2015zsa} 
  E.~Hijano, P.~Kraus, E.~Perlmutter and R.~Snively,
  ``Witten Diagrams Revisited: The AdS Geometry of Conformal Blocks,''
  JHEP {\bf 1601}, 146 (2016)
  doi:10.1007/JHEP01(2016)146
  [arXiv:1508.00501 [hep-th]].
  %\cite{Hijano:2015qja}
  
\bibitem{Hijano:2015qja} 
  E.~Hijano, P.~Kraus, E.~Perlmutter and R.~Snively,
  ``Semiclassical Virasoro blocks from AdS$_{3}$ gravity,''
  JHEP {\bf 1512}, 077 (2015)
  doi:10.1007/JHEP12(2015)077
  [arXiv:1508.04987 [hep-th]].
  %%CITATION = doi:10.1007/JHEP12(2015)077;%%
  
  

  
  
\bibitem{Casini:2011kv} 
  H.~Casini, M.~Huerta and R.~C.~Myers,
  ``Towards a derivation of holographic entanglement entropy,''
  JHEP {\bf 1105}, 036 (2011)
  doi:10.1007/JHEP05(2011)036
  [arXiv:1102.0440 [hep-th]].
%\bibitem{16}
%Banks, Tom, et al. "AdS dynamics from conformal field theory." arXiv preprint hep-th/9808016 (1998).
\bibitem{Hamilton:2006az} 
  A.~Hamilton, D.~N.~Kabat, G.~Lifschytz and D.~A.~Lowe,
  ``Holographic representation of local bulk operators,''
  Phys.\ Rev.\ D {\bf 74}, 066009 (2006)
  doi:10.1103/PhysRevD.74.066009
  [hep-th/0606141].
\bibitem{Hamilton:2005ju} 
  A.~Hamilton, D.~N.~Kabat, G.~Lifschytz and D.~A.~Lowe,
  ``Local bulk operators in AdS/CFT: A Boundary view of horizons and locality,''
  Phys.\ Rev.\ D {\bf 73}, 086003 (2006)
  doi:10.1103/PhysRevD.73.086003
  [hep-th/0506118].
\bibitem{Guica:2016pid} 
  M.~Guica,
  ``Bulk fields from the boundary OPE,''
  arXiv:1610.08952 [hep-th].
\bibitem{18.1}
ymsc.tsinghua.edu.cn:8090/strings/slides/parallel/4/Sully%20v2.pptx
%\bibitem{18.2}
%Guica, Monica. "Bulk fields from the boundary OPE." arXiv preprint arXiv:1610.08952 (2016).
\bibitem{Czech:2016tqr} 
  B.~Czech, L.~Lamprou, S.~McCandlish, B.~Mosk and J.~Sully,
  ``Equivalent Equations of Motion for Gravity and Entropy,''
  JHEP {\bf 1702}, 004 (2017)
  doi:10.1007/JHEP02(2017)004
  [arXiv:1608.06282 [hep-th]].
\bibitem{Lashkari:2013koa} 
  N.~Lashkari, M.~B.~McDermott and M.~Van Raamsdonk,
  ``Gravitational dynamics from entanglement 'thermodynamics',''
  JHEP {\bf 1404}, 195 (2014)
  doi:10.1007/JHEP04(2014)195
  [arXiv:1308.3716 [hep-th]].
\bibitem{Faulkner:2013ica} 
  T.~Faulkner, M.~Guica, T.~Hartman, R.~C.~Myers and M.~Van Raamsdonk,
  ``Gravitation from Entanglement in Holographic CFTs,''
  JHEP {\bf 1403}, 051 (2014)
  doi:10.1007/JHEP03(2014)051
  [arXiv:1312.7856 [hep-th]].
\bibitem{Swingle:2014uza} 
  B.~Swingle and M.~Van Raamsdonk,
  ``Universality of Gravity from Entanglement,''
  arXiv:1405.2933 [hep-th].
\bibitem{VanRaamsdonk:2016exw} 
  M.~Van Raamsdonk,
  ``Lectures on Gravity and Entanglement,''
  arXiv:1609.00026 [hep-th].
\bibitem{Balasubramanian:2013lsa} 
  V.~Balasubramanian, B.~D.~Chowdhury, B.~Czech, J.~de Boer and M.~P.~Heller,
  ``Bulk curves from boundary data in holography,''
  Phys.\ Rev.\ D {\bf 89}, no. 8, 086004 (2014)
  doi:10.1103/PhysRevD.89.086004
  [arXiv:1310.4204 [hep-th]].
 \bibitem{Myers:2014jia} 
  R.~C.~Myers, J.~Rao and S.~Sugishita,
  ``Holographic Holes in Higher Dimensions,''
  JHEP {\bf 1406}, 044 (2014)
  doi:10.1007/JHEP06(2014)044
  [arXiv:1403.3416 [hep-th]].
  \bibitem{Headrick:2014eia} 
  M.~Headrick, R.~C.~Myers and J.~Wien,
  ``Holographic Holes and Differential Entropy,''
  JHEP {\bf 1410}, 149 (2014)
  doi:10.1007/JHEP10(2014)149
  [arXiv:1408.4770 [hep-th]]. 
\bibitem{Czech:2014wka} 
  B.~Czech, X.~Dong and J.~Sully,
  ``Holographic Reconstruction of General Bulk Surfaces,''
  JHEP {\bf 1411}, 015 (2014)
  doi:10.1007/JHEP11(2014)015
  [arXiv:1406.4889 [hep-th]].
\bibitem{Czech:2014ppa} 
  B.~Czech and L.~Lamprou,
  ``Holographic definition of points and distances,''
  Phys.\ Rev.\ D {\bf 90}, 106005 (2014)
  doi:10.1103/PhysRevD.90.106005
  [arXiv:1409.4473 [hep-th]].

%\cite{Asplund:2016koz}
\bibitem{Asplund:2016koz} 
  C.~T.~Asplund, N.~Callebaut and C.~Zukowski,
  ``Equivalence of Emergent de Sitter Spaces from Conformal Field Theory,''
  JHEP {\bf 1609}, 154 (2016)
  doi:10.1007/JHEP09(2016)154
  [arXiv:1604.02687 [hep-th]].

%\bibitem{50}
%Asplund, Curtis T., Nele Callebaut, and Claire Zukowski. "Equivalence of emergent de Sitter spaces from conformal field theory." %Journal of High Energy Physics 2016.9 (2016): 154.
\bibitem{Calabrese:2004eu} 
  P.~Calabrese and J.~L.~Cardy,
  ``Entanglement entropy and quantum field theory,''
  J.\ Stat.\ Mech.\  {\bf 0406}, P06002 (2004)
  doi:10.1088/1742-5468/2004/06/P06002
  [hep-th/0405152].
\bibitem{Calabrese:2009qy} 
  P.~Calabrese and J.~Cardy,
  ``Entanglement entropy and conformal field theory,''
  J.\ Phys.\ A {\bf 42}, 504005 (2009)
  doi:10.1088/1751-8113/42/50/504005
  [arXiv:0905.4013 [cond-mat.stat-mech]].
\bibitem{Calabrese:2007mtj} 
  P.~Calabrese and J.~Cardy,
  ``Entanglement and correlation functions following a local quench: a conformal field theory approach,''
  J.\ Stat.\ Mech.\  {\bf 0710}, no. 10, P10004 (2007)
  doi:10.1088/1742-5468/2007/10/P10004
%\cite{deBoer:2015kda}
\bibitem{deBoer:2015kda}
  J.~de Boer, M.~P.~Heller, R.~C.~Myers and Y.~Neiman,
  ``Holographic de Sitter Geometry from Entanglement in Conformal Field Theory,''
  Phys.\ Rev.\ Lett.\  {\bf 116} (2016) no.6,  061602
  doi:10.1103/PhysRevLett.116.061602
  [arXiv:1509.00113 [hep-th]].

%\cite{Cresswell:2017mbk}
\bibitem{Cresswell:2017mbk} 
  J.~C.~Cresswell and A.~W.~Peet,
  ``Kinematic space for conical defects,''
  JHEP {\bf 1711}, 155 (2017)
  doi:10.1007/JHEP11(2017)155
  [arXiv:1708.09838 [hep-th]].
  
%\cite{Takayanagi:2011zk}
%\bibitem{Takayanagi:2011zk} 
 % T.~Takayanagi,
  %``Holographic Dual of BCFT,''
  %Phys.\ Rev.\ Lett.\  {\bf 107}, 101602 (2011)
  %doi:10.1103/PhysRevLett.107.101602
  %[arXiv:1105.5165 [hep-th]].
  %%CITATION = doi:10.1103/PhysRevLett.107.101602;%%
%\cite{Ryu:2006bv}
%\cite{Fujita:2011fp}
%\bibitem{A} 
  %M.~Fujita, T.~Takayanagii and E.~Tonni,
  %``Aspects of AdS/BCFT,''
  %JHEP {\bf 1111}, 043 (2011)
  %doi:10.1007/JHEP11(2011)043
  %[arXiv:1108.5152 [hep-th]].
  %%CITATION = doi:10.1007/JHEP11(2011)043;%%
%\bibitem{Ryu:2006bv} 
%  S.~Ryu and T.~Takayanagi,
%  ``Holographic derivation of entanglement entropy from AdS/CFT,''
%  Phys.\ Rev.\ Lett.\  {\bf 96}, 181602 (2006)
%  doi:10.1103/PhysRevLett.96.181602
%  [hep-th/0603001].
  %%CITATION = doi:10.1103/PhysRevLett.96.181602;%%
%\cite{Fujita:2011fp}
%\bibitem{Fujita:2011fp} 
 % M.~Fujita, T.~Takayanagi and E.~Tonni,
  %``Aspects of AdS/BCFT,''
  %JHEP {\bf 1111}, 043 (2011)
  %doi:10.1007/JHEP11(2011)043
  %[arXiv:1108.5152 [hep-th]].
  %%CITATION = doi:10.1007/JHEP11(2011)043;%% 
  %\cite{Alishahiha:2011rg}
%\bibitem{A}
%\bibitem{Alishahiha:2011rg} 
  %M.~Alishahiha and R.~Fareghbal,
  %``Boundary CFT from Holography,''
  %Phys.\ Rev.\ D {\bf 84}, 106002 (2011)
  %doi:10.1103/PhysRevD.84.106002
  %[arXiv:1108.5607 [hep-th]].
  %%CITATION = doi:10.1103/PhysRevD.84.106002;%%
%\cite{Chiodaroli:2011fn}
%\bibitem{Chiodaroli:2011fn} 
 % M.~Chiodaroli, E.~D'Hoker and M.~Gutperle,
 % ``Simple Holographic Duals to Boundary CFTs,''
  %JHEP {\bf 1202}, 005 (2012)
  %doi:10.1007/JHEP02(2012)005
  %[arXiv:1111.6912 [hep-th]].
  %%CITATION = doi:10.1007/JHEP02(2012)005;%%  
  %\cite{Setare:2012ks}
%\bibitem{Setare:2012ks} 
  %M.~R.~Setare and V.~Kamali,
  %``Anti--de Sitter/ boundary conformal field theory correspondence in the nonrelativistic limit,''
  %Eur.\ Phys.\ J.\ C {\bf 72}, 2115 (2012)
  %doi:10.1140/epjc/s10052-012-2115-x
  %[arXiv:1202.4917 [hep-th]].
  %%CITATION = doi:10.1140/epjc/s10052-012-2115-x;%%
  %\cite{Fujita:2012fp}
%\bibitem{Fujita:2012fp} 
  %M.~Fujita, M.~Kaminski and A.~Karch,
  %``SL(2,Z) Action on AdS/BCFT and Hall Conductivities,''
 % JHEP {\bf 1207}, 150 (2012)
 % doi:10.1007/JHEP07(2012)150
  %[arXiv:1204.0012 [hep-th]].
  %%CITATION = doi:10.1007/JHEP07(2012)150;%%
%\cite{GarciaGarcia:2012zd}
%\bibitem{GarciaGarcia:2012zd} 
  %A.~M.~Garcia-Garcia, J.~E.~Santos and B.~Way,
  %``Holographic Description of Finite Size Effects in Strongly Coupled Superconductors,''
  %Phys.\ Rev.\ B {\bf 86}, 064526 (2012)
  %doi:10.1103/PhysRevB.86.064526
  %[arXiv:1204.4189 [hep-th]].
  %%CITATION = doi:10.1103/PhysRevB.86.064526;%%  
  %\cite{Nozaki:2012qd}
%\bibitem{Nozaki:2012qd} 
 % M.~Nozaki, T.~Takayanagi and T.~Ugajin,
  %``Central Charges for BCFTs and Holography,''
 % JHEP {\bf 1206}, 066 (2012)
 % doi:10.1007/JHEP06(2012)066
 % [arXiv:1205.1573 [hep-th]].
  %%CITATION = doi:10.1007/JHEP06(2012)066;%%
  %\cite{Chiodaroli:2012vc}
%\bibitem{Chiodaroli:2012vc} 
  %M.~Chiodaroli, E.~D'Hoker and M.~Gutperle,
 % ``Holographic duals of Boundary CFTs,''
%  JHEP {\bf 1207}, 177 (2012)
  %doi:10.1007/JHEP07(2012)177
 % [arXiv:1205.5303 [hep-th]].
  %%CITATION = doi:10.1007/JHEP07(2012)177;%%
  %\cite{Gutperle:2012hy}
%\bibitem{Gutperle:2012hy} 
%  M.~Gutperle and J.~Samani,
  %``Holographic RG-flows and Boundary CFTs,''
 % Phys.\ Rev.\ D {\bf 86}, 106007 (2012)
%  doi:10.1103/PhysRevD.86.106007
 % [arXiv:1207.7325 [hep-th]].
  %%CITATION = doi:10.1103/PhysRevD.86.106007;%%
  %\cite{Melnikov:2012tb}
%\bibitem{Melnikov:2012tb} 
  %D.~Melnikov, E.~Orazi and P.~Sodano,
  %``On the AdS/BCFT Approach to Quantum Hall Systems,''
 % JHEP {\bf 1305}, 116 (2013)
 % doi:10.1007/JHEP05(2013)116
  %[arXiv:1211.1416 [hep-th]].
  %%CITATION = doi:10.1007/JHEP05(2013)116;%%
  %\cite{Jensen:2013lxa}
%\bibitem{Jensen:2013lxa} 
 % K.~Jensen and A.~O'Bannon,
  %``Holography, Entanglement Entropy, and Conformal Field Theories with Boundaries or Defects,''
  %Phys.\ Rev.\ D {\bf 88}, no. 10, 106006 (2013)
 % doi:10.1103/PhysRevD.88.106006
 % [arXiv:1309.4523 [hep-th]].
  %%CITATION = doi:10.1103/PhysRevD.88.106006;%%
  %\cite{Korovin:2013gha}
%\bibitem{Korovin:2013gha} 
%  Y.~Korovin,
%  ``First order formalism for the holographic duals of defect CFTs,''
 % JHEP {\bf 1404}, 152 (2014)
 % doi:10.1007/JHEP04(2014)152
 % [arXiv:1312.0089 [hep-th]].
  %%CITATION = doi:10.1007/JHEP04(2014)152;%%
  %\cite{Estes:2014hka}
%\bibitem{Estes:2014hka} 
 % J.~Estes, K.~Jensen, A.~O'Bannon, E.~Tsatis and T.~Wrase,
%  ``On Holographic Defect Entropy,''
%  JHEP {\bf 1405}, 084 (2014)
%  doi:10.1007/JHEP05(2014)084
%  [arXiv:1403.6475 [hep-th]].
  %%CITATION = doi:10.1007/JHEP05(2014)084;%%
  %\cite{Astaneh:2014fga}
%\bibitem{Astaneh:2014fga} 
 % A.~F.~Astaneh and A.~E.~Mosaffa,
%  ``Quantum Local Quench, AdS/BCFT and Yo-Yo String,''
%  JHEP {\bf 1505}, 107 (2015)
%  doi:10.1007/JHEP05(2015)107
%  [arXiv:1405.5469 [hep-th]].
  %%CITATION = doi:10.1007/JHEP05(2015)107;%%
  %\cite{Magan:2014dwa}
%\bibitem{Magan:2014dwa} 
 % J.~M.~Magán, D.~Melnikov and M.~R.~O.~Silva,
  %``Black Holes in AdS/BCFT and Fluid/Gravity Correspondence,''
 % JHEP {\bf 1411}, 069 (2014)
 % doi:10.1007/JHEP11(2014)069
 % [arXiv:1408.2580 [hep-th]].
  %%CITATION = doi:10.1007/JHEP11(2014)069;%%
  %\cite{Najian:2014waa}
%\bibitem{Najian:2014waa} 
 % B.~Najian,
%  ``AdS/BCFT correspondence, holographic g-theorem and Gauss-Bonnet gravity,''
%  Int.\ J.\ Mod.\ Phys.\ A {\bf 29}, no. 24, 1450139 (2014).
%  doi:10.1142/S0217751X14501395
  %%CITATION = doi:10.1142/S0217751X14501395;%%
  %\cite{Miyaji:2014mca}
%\bibitem{Miyaji:2014mca} 
%  M.~Miyaji, S.~Ryu, T.~Takayanagi and X.~Wen,
 % ``Boundary States as Holographic Duals of Trivial Spacetimes,''
%  JHEP {\bf 1505}, 152 (2015)
 % doi:10.1007/JHEP05(2015)152
 % [arXiv:1412.6226 [hep-th]].
  %%CITATION = doi:10.1007/JHEP05(2015)152;%%
  %\cite{Estes:2015jha}
%\bibitem{Estes:2015jha} 
 % J.~Estes,
%  ``Finite temperature holographic duals of 2-dimensional BCFTs,''
 % JHEP {\bf 1507}, 020 (2015)
 % doi:10.1007/JHEP07(2015)020
%  [arXiv:1503.07375 [hep-th]].
  %%CITATION = doi:10.1007/JHEP07(2015)020;%%
  %\cite{Flory:2017ftd}
%\bibitem{Flory:2017ftd} 
 % M.~Flory,
 % ``A complexity/fidelity susceptibility g-theorem for AdS$_3$/BCFT$_2$,''
%  arXiv:1702.06386 [hep-th].
%\bibitem{Astaneh:2017ghi} 
 % A.~F.~Astaneh and S.~N.~Solodukhin,
%  ``Holographic calculation of boundary terms in conformal anomaly,''
%  arXiv:1702.00566 [hep-th].
  %%CITATION = ARXIV:1702.00566;%%
  %\cite{Miao:2017gyt}
%\bibitem{Miao:2017gyt} 
%  R.~X.~Miao, C.~S.~Chu and W.~Z.~Guo,
%  ``A New Proposal for Holographic BCFT,''
%  arXiv:1701.04275 [hep-th].
  %%CITATION = ARXIV:1701.04275;%%
  %\cite{Chu:2017aab}
%\bibitem{Chu:2017aab} 
 % C.~S.~Chu, R.~X.~Miao and W.~Z.~Guo,
 % ``On New Proposal for Holographic BCFT,''
%  arXiv:1701.07202 [hep-th].
  %%CITATION = ARXIV:1701.07202;%%
  %\cite{Gutperle:2016gfe}
%\bibitem{Gutperle:2016gfe} 
%  M.~Gutperle and A.~Trivella,
%  ``A note on entanglement entropy and regularization in holographic interface theories,''
%  arXiv:1611.07595 [hep-th].
  %%CITATION = ARXIV:1611.07595;%%
  %\cite{Flory:2016mas}
%\bibitem{Flory:2016mas} 
%  M.~Flory,
%  ``Entanglement and defect entropies in gauge/gravity duality,''
  %\cite{Mandal:2016cdw}
%\bibitem{Mandal:2016cdw} 
%  G.~Mandal, R.~Sinha and T.~Ugajin,
 % ``Finite size effect on dynamical entanglement entropy: CFT and holography,''
%  arXiv:1604.07830 [hep-th].
%\bibitem{47}
%Takayanagi, Tadashi. "Holographic dual of a boundary conformal field theory." Physical review letters 107.10 (2011): 101602.
%\bibitem{48}
%Fujita, Mitsutoshi, Tadashi Takayanagi, and Erik Tonni. "Aspects of AdS/BCFT." Journal of High Energy Physics 2011.11 (2011): 1-40.
%\bibitem{49}
%Miao, Rong-Xin, Chong-Sun Chu, and Wu-Zhong Guo. "A New Proposal for Holographic BCFT." arXiv preprint arXiv:1701.04275 (2017).
%\bibitem{40}
%Astaneh, Amin Faraji, and Amir Esmaeil Mosaffa. "Quantum local quench, AdS/BCFT and yo-yo string." arXiv preprint arXiv:1405.5469 (2014).
%\bibitem{Verlinde:2015qfa} 
 % H.~Verlinde,
%  ``Poking Holes in AdS/CFT: Bulk Fields from Boundary States,''
 % arXiv:1505.05069 [hep-th].
%\bibitem{Nakayama:2015mva} 
%  Y.~Nakayama and H.~Ooguri,
%  ``Bulk Locality and Boundary Creating Operators,''
%  JHEP {\bf 1510}, 114 (2015)
%  doi:10.1007/JHEP10(2015)114
%  [arXiv:1507.04130 [hep-th]].
%\bibitem{Nakayama:2016xvw} 
%  Y.~Nakayama and H.~Ooguri,
%  ``Bulk Local States and Crosscaps in Holographic CFT,''
%  JHEP {\bf 1610}, 085 (2016)
%  doi:10.1007/JHEP10(2016)085
%  [arXiv:1605.00334 [hep-th]].
%\bibitem{Lewkowycz:2016ukf} 
 % A.~Lewkowycz, G.~J.~Turiaci and H.~Verlinde,
 % ``A CFT Perspective on Gravitational Dressing and Bulk Locality,''
%  JHEP {\bf 1701}, 004 (2017)
%  doi:10.1007/JHEP01(2017)004
%  [arXiv:1608.08977 [hep-th]].
 % \bibitem{Kabat:2017mun} 
 % D.~Kabat and G.~Lifschytz,
%  ``Local bulk physics from intersecting modular Hamiltonians,''
%  JHEP {\bf 1706}, 120 (2017)
 % doi:10.1007/JHEP06(2017)120
%  [arXiv:1703.06523 [hep-th]].
\bibitem{Czech:2016nxc} 
  B.~Czech, P.~H.~Nguyen and S.~Swaminathan,
  ``A defect in holographic interpretations of tensor networks,''
  JHEP {\bf 1703}, 090 (2017)
  doi:10.1007/JHEP03(2017)090
  [arXiv:1612.05698 [hep-th]]..

%\cite{deBoer:2016pqk}
\bibitem{deBoer:2016pqk} 
  J.~de Boer, F.~M.~Haehl, M.~P.~Heller and R.~C.~Myers,
  ``Entanglement, holography and causal diamonds,''
  JHEP {\bf 1608}, 162 (2016)
  doi:10.1007/JHEP08(2016)162
  [arXiv:1606.03307 [hep-th]].
  %%CITATION = doi:10.1007/JHEP08(2016)162;%%

%\cite{Holzhey:1994we}
\bibitem{Holzhey:1994we} 
  C.~Holzhey, F.~Larsen and F.~Wilczek,
  ``Geometric and renormalized entropy in conformal field theory,''
  Nucl.\ Phys.\ B {\bf 424}, 443 (1994)
  doi:10.1016/0550-3213(94)90402-2
  [hep-th/9403108].
  %%CITATION = doi:10.1016/0550-3213(94)90402-2;%%

%\cite{Karch:2017fuh}
\bibitem{Karch:2017fuh} 
  A.~Karch, J.~Sully, C.~F.~Uhlemann and D.~G.~E.~Walker,
  ``Boundary Kinematic Space,''
  JHEP {\bf 1708}, 039 (2017)
  doi:10.1007/JHEP08(2017)039
  [arXiv:1703.02990 [hep-th]].
  
  \bibitem{Sheikh-Jabbari:2016znt} 
  M.~M.~Sheikh-Jabbari and H.~Yavartanoo,
  ``Excitation entanglement entropy in two dimensional conformal field theories,''
  Phys.\ Rev.\ D {\bf 94}, no. 12, 126006 (2016)
  doi:10.1103/PhysRevD.94.126006
  [arXiv:1605.00341 [hep-th]].
  
  \bibitem{Rosenhaus:2014zza} 
  V.~Rosenhaus and M.~Smolkin,
  ``Entanglement Entropy for Relevant and Geometric Perturbations,''
  JHEP {\bf 1502}, 015 (2015)
  doi:10.1007/JHEP02(2015)015
  [arXiv:1410.6530 [hep-th]].
  
  \bibitem{Rosenhaus:2014ula} 
  V.~Rosenhaus and M.~Smolkin,
  ``Entanglement entropy, planar surfaces, and spectral functions,''
  JHEP {\bf 1409}, 119 (2014)
  doi:10.1007/JHEP09(2014)119
  [arXiv:1407.2891 [hep-th]].
  
  \bibitem{Rosenhaus:2014nha} 
  V.~Rosenhaus and M.~Smolkin,
  ``Entanglement Entropy Flow and the Ward Identity,''
  Phys.\ Rev.\ Lett.\  {\bf 113}, no. 26, 261602 (2014)
  doi:10.1103/PhysRevLett.113.261602
  [arXiv:1406.2716 [hep-th]].
  
  \bibitem{Rosenhaus:2014woa} 
  V.~Rosenhaus and M.~Smolkin,
  ``Entanglement Entropy: A Perturbative Calculation,''
  JHEP {\bf 1412}, 179 (2014)
  doi:10.1007/JHEP12(2014)179
  [arXiv:1403.3733 [hep-th]].
  
 
   %%CITATION = ARXIV:1806.02289;%%
  \bibitem{Faulkner:2014jva} 
  T.~Faulkner,
  ``Bulk Emergence and the RG Flow of Entanglement Entropy,''
  JHEP {\bf 1505}, 033 (2015)
  doi:10.1007/JHEP05(2015)033
  [arXiv:1412.5648 [hep-th]].
  
   %\cite{Antonelli:2018qwz}
\bibitem{Antonelli:2018qwz} 
  R.~Antonelli, I.~Basile and A.~Bombini,
  ``AdS Vacuum Bubbles, Holography and dual RG Flows,''
  arXiv:1806.02289 [hep-th].
  
  \bibitem{Casini:2011kv} 
  H.~Casini, M.~Huerta and R.~C.~Myers,
  ``Towards a derivation of holographic entanglement entropy,''
  JHEP {\bf 1105}, 036 (2011)
  doi:10.1007/JHEP05(2011)036
  [arXiv:1102.0440 [hep-th]].
  
  \bibitem{Cardy:2016fqc} 
  J.~Cardy and E.~Tonni,
  ``Entanglement hamiltonians in two-dimensional conformal field theory,''
  J.\ Stat.\ Mech.\  {\bf 1612}, no. 12, 123103 (2016)
  doi:10.1088/1742-5468/2016/12/123103
  [arXiv:1608.01283 [cond-mat.stat-mech]].
%\bibitem{sully}
%Karch, Andreas, James Sully, Christoph F. Uhlemann, and Devin GE Walker. "Boundary Kinematic Space." arXiv preprint arXiv:1703.02990 (2017).
%=======================================================
%\cite{Calabrese:2004eu}
%\bibitem{Calabrese:2004eu}
%  P.~Calabrese and J.~L.~Cardy,
  %``Entanglement entropy and quantum field theory,''
%  J.\ Stat.\ Mech.\  {\bf 0406} (2004) P06002
%  doi:10.1088/1742-5468/2004/06/P06002
%  [hep-th/0405152].
%  %CITATION = doi:10.1088/1742-5468/2004/06/P06002;%%
%  608 citations counted in INSPIRE as of 21 Feb 2017
  %CITATION = doi:10.1007/JHEP08(2016)162;%%
%  15 citations counted in INSPIRE as of 13 Feb 2017
  
%\cite{Czech:2014wka}
%\bibitem{Czech:2014wka}
%  B.~Czech, X.~Dong and J.~Sully,
  %``Holographic Reconstruction of General Bulk Surfaces,''
%  JHEP {\bf 1411} (2014) 015
%  doi:10.1007/JHEP11(2014)015
%  [arXiv:1406.4889 [hep-th]].
  %%CITATION = doi:10.1007/JHEP11(2014)015;%%
  %33 citations counted in INSPIRE as of 11 Feb 2017
  
%  J.\ Stat.\ Mech.\  {\bf 0406} (2004) P06002
%  doi:10.1088/1742-5468/2004/06/P06002
%  [hep-th/0405152].
  %%CITATION = doi:10.1088/1742-5468/2004/06/P06002;%%
  %608 citations counted in INSPIRE as of 21 Feb 2017
  
  
%\cite{Calabrese:2007mtj}
%\bibitem{Calabrese:2007mtj}
%  P.~Calabrese and J.~Cardy,
  %``Entanglement and correlation functions following a local quench: a conformal field theory approach,''
%  J.\ Stat.\ Mech.\  {\bf 0710} (2007) no.10,  P10004
%  doi:10.1088/1742-5468/2007/10/P10004
%  [arXiv:0708.3750 [quant-ph]].
  %%CITATION = doi:10.1088/1742-5468/2007/10/P10004;%%
  %50 citations counted in INSPIRE as of 21 Feb 2017
  
%\cite{Astaneh:2014fga}
%\bibitem{Astaneh:2014fga}
%  A.~F.~Astaneh and A.~E.~Mosaffa,
  %``Quantum Local Quench, AdS/BCFT and Yo-Yo String,''
%  JHEP {\bf 1505} (2015) 107
%  doi:10.1007/JHEP05(2015)107
%  [arXiv:1405.5469 [hep-th]].
  %%CITATION = doi:10.1007/JHEP05(2015)107;%%
  %11 citations counted in INSPIRE as of 21 Feb 2017
    
   %\cite{Czech:2014wka}
%\bibitem{Czech:2014wka}
%  B.~Czech, X.~Dong and J.~Sully,
  %``Holographic Reconstruction of General Bulk Surfaces,''
%  JHEP {\bf 1411} (2014) 015
%  doi:10.1007/JHEP11(2014)015
%  [arXiv:1406.4889 [hep-th]].
  %%CITATION = doi:10.1007/JHEP11(2014)015;%%
  %33 citations counted in INSPIRE as of 11 Feb 2017

  %=============================
%\bibitem{19}
%Heemskerk, Idse, et al. "Holography from conformal field theory." Journal of High Energy Physics 2009.10 (2009): 079.
%\bibitem{bh1}
%Bombelli, Luca, et al. "Quantum source of entropy for black holes." Physical Review D 34.2 (1986): 373.
%\bibitem{bh2}
%Srednicki, Mark. "Entropy and area." Physical Review Letters 71.5 (1993): 666.
%\bibitem{bh3}
%Holzhey, Christoph, Finn Larsen, and Frank Wilczek. "Geometric and renormalized entropy in conformal field theory." Nuclear Physics B 424.3 (1994): 443-467.
%\bibitem{8}
%Calabrese, Pasquale, and John Cardy. "Entanglement entropy and quantum field theory." Journal of Statistical Mechanics: Theory and Experiment 2004.06 (2004): P06002.
%\bibitem{9}
%Calabrese, Pasquale, and John Cardy. "Entanglement entropy and conformal field theory." Journal of Physics A: Mathematical and Theoretical 42.50 (2009): 504005.
%\bibitem{10}
%Casini, H., and M. Huerta. "Entanglement entropy in free quantum field theory." Journal of Physics A: Mathematical and Theoretical 42.50 (2009): 504007.
%\bibitem{11}
%Ryu, Shinsei, and Tadashi Takayanagi. "Holographic derivation of entanglement entropy from the anti–de sitter space/conformal field theory correspondence." Physical review letters 96.18 (2006): 181602.
%\bibitem{12}
%Ryu, Shinsei, and Tadashi Takayanagi. "Aspects of holographic entanglement entropy." Journal of High Energy Physics 2006.08 (2006): 045.
%\bibitem{13}
%Hubeny, Veronika E., Mukund Rangamani, and Tadashi Takayanagi. "A covariant holographic entanglement entropy proposal." Journal of High Energy Physics 2007.07 (2007): 062.
%\bibitem{14}
%Nishioka, Tatsuma, Shinsei Ryu, and Tadashi Takayanagi. "Holographic entanglement entropy: an overview." Journal of Physics A: Mathematical and Theoretical 42.50 (2009): 504008.
%\bibitem{15}
%Headrick, Matthew, and Tadashi Takayanagi. "Holographic proof of the strong subadditivity of entanglement entropy." Physical Review D 76.10 (2007): 106013.
%\bibitem{15.1}
%Hayden, Patrick, Matthew Headrick, and Alexander Maloney. "Holographic mutual information is monogamous." Physical Review D 87.4 (2013): 046003.
%
%\bibitem{24}
%Lin, Jennifer, et al. "Tomography from entanglement." arXiv preprint arXiv:1412.1879 (2014).
%\bibitem{25}
%Lashkari, Nima, et al. "Inviolable energy conditions from entanglement inequalities." arXiv preprint arXiv:1412.3514 (2014).
%\bibitem{26}
%Van Raamsdonk, Mark. "Comments on quantum gravity and entanglement." arXiv preprint arXiv:0907.2939 (2009).
%\bibitem{27}
%Van Raamsdonk, Mark. "Building up spacetime with quantum entanglement." General Relativity and Gravitation 42.10 (2010): 2323-2329.
%\bibitem{28}
%Maldacena, Juan, and Leonard Susskind. "Cool horizons for entangled black holes." Fortschritte der Physik 61.9 (2013): 781-811.
%
%\bibitem{34}
%Czech, Bartlomiej, et al. "Integral geometry and holography." arXiv preprint arXiv:1505.05515 (2015).
%\bibitem{35} 
%Czech, Bartlomiej, et al. "A Stereoscopic Look into the Bulk." arXiv preprint arXiv:1604.03110 (2016).
%\cite{Czech:2016xec}
%\bibitem{46}
%de Boer, Jan, et al. "Entanglement, holography and causal diamonds." arXiv preprint arXiv:1606.03307 (2016).
%
%\bibitem{37}

%
%%\bibitem{41}
%Anninos, Dionysios, Gim Seng Ng, and Andrew Strominger. "Future boundary conditions in de Sitter space." Journal of High Energy Physics 2012.2 (2012): 1-17.
%
%
%\bibitem{42}
%Zamolodchikov, Alexander B. "Irreversibility of the Flux of the Renormalization Group in a 2D Field Theory." JETP lett 43.12 (1986): 730-732.
%\bibitem{43}
%Casini, H., and M. Huerta. "A finite entanglement entropy and the c-theorem." Physics Letters B 600.1 (2004): 142-150.
%\bibitem{44}
%Casini, Horacio, and Marina Huerta. "A c-theorem for entanglement entropy." Journal of Physics A: Mathematical and Theoretical 40.25 (2007): 7031.
%\bibitem{45}
%Casini, H., and Marina Huerta. "Renormalization group running of the entanglement entropy of a circle." Physical Review D 85.12 (2012): 125016.
\end{thebibliography}
\end{document}